\documentclass[10pt,aps,prx,twocolumn,superscriptaddress]{revtex4-2}
\usepackage[colorlinks,allcolors=blue]{hyperref}
\usepackage{amsmath,amsfonts,tikz}
\usepackage{graphicx,xcolor,multirow}

\def\be{\begin{equation}}
\def\ee{\end{equation}}
\def\bea{\begin{eqnarray}}
\def\eea{\end{eqnarray}}
\def\nn{\nonumber}

\usepackage{lineno}

\newcommand{\ket}[1]{\left|#1\right\rangle}
\newcommand{\bra}[1]{\left\langle #1\right|}
\newcommand{\braket}[1]{\left\langle #1 \right\rangle}

\newcommand{\rbraket}[1]{\left( #1 \right)} 
\newcommand{\sbraket}[1]{\left[ #1 \right]} 
\newcommand{\Tr}[1]{\operatorname{\textnormal{Tr}}\left( {#1} \right)}

\begin{document}
\title{Quantum Sensing with Topological-Paired Bound States }

\author{Tao Zhang}
\email{These authors contribute equally to this work}
\affiliation{Department of Physics and State Key Laboratory of Low Dimensional Quantum Physics, Tsinghua University, Beijing, 100084, China}

\author{Peng Xu}
\email{These authors contribute equally to this work}
\affiliation{School of Physics, Zhengzhou University, Zhengzhou 450001, China}
\affiliation{Institute of Quantum Materials and Physics, Henan Academy of Sciences, Zhengzhou 450046, China}

\author{Jiazhong Hu}
\email{hujiazhong01@ultracold.cn}
\affiliation{Department of Physics and State Key Laboratory of Low Dimensional Quantum Physics, Tsinghua University, Beijing, 100084, China}
\affiliation{Frontier Science Center for Quantum Information and Collaborative Innovation Center of Quantum Matter, Beijing, 100084, China}

\author{Xingze Qiu}
\email{xingze@tongji.edu.cn}
\affiliation{School of Physics Science and Engineering, Tongji University, Shanghai 200092, China}

\date{\today}

\begin{abstract}
We present an efficient and robust protocol for quantum-enhanced sensing using a single qubit in the topological waveguide system. 
Our method relies on the topological-paired bound states, which are localized near the qubit and can be effectively regarded as a two-level system. 
Through the lens of Bayesian inference theory, we show that the sensitivity can reach the Heisenberg limit across a large field range. 
Inheriting from the topological robustness of the waveguide, our sensing protocol is robust against local perturbations. 
Besides, our sensing protocol utilizes a product state as the initial state, which can be easily prepared in experiments. 
We expect this approach would pave the way toward robust topological quantum sensors based on near-term quantum platforms such as superconducting qubits and Rydberg arrays. 
\end{abstract}

\maketitle

\section{Introduction}

Quantum sensing (QS) exploits quantum advantages for increasing the sensitivity of estimating certain physical parameters, and represents a key technology in both fundamental science and concrete applications \cite{DRC2017RMP}. 
The sensitivity of QS can surpass the shot-noise limit of classical sensors and even reach the Heisenberg limit, which is imposed by quantum mechanics \cite{Lloyd2004Science, Lloyd2006PRL}. 
Proposed applications of QS include magnetometry \cite{Lukin2008NP, Tanaka2015PRL}, electrometry \cite{Dolde2011NP, Facon2016Nature}, gravitational wave detection \cite{Caves1981PRD, Tse2019PRL}, and dark matter detection \cite{Rajendran2017PRD}. 

\begin{figure}[h!]
\centering
\includegraphics[width=0.48\textwidth]{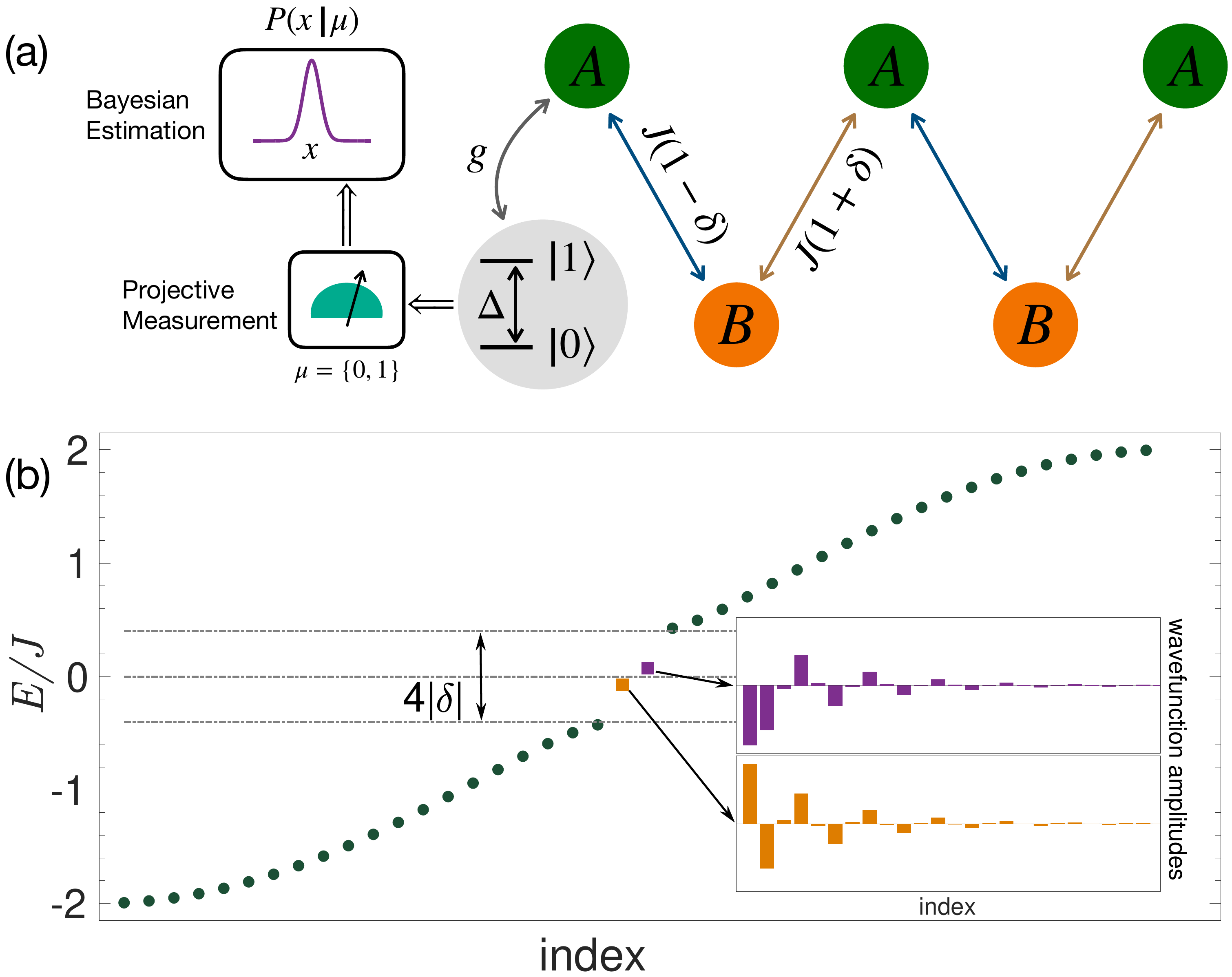}
\caption{
(a) Schematic illustration of quantum sensing with topological-paired bound states. 
The SSH chain is characterized by having alternating hopping amplitudes $J(1\pm\delta)$ between sublattices $A$ and $B$.  
A two-level QE with transition frequency $\Delta$ couples to sublattice $A$ at the leftmost site with a coupling strength $g$. 
The initial state of the system is a product state $\ket{\psi_0}=\ket{1}\otimes\ket{\rm vac}$, i.e., the QE and the SSH chain lie in the excited state $\ket{1}$ and the vacuum state $\ket{\rm vac}$, respectively. 
After a sensing time, the state of the QE is measured, and the outcome $\mu=\{0,1\}$ is used to estimate the unknown parameter $x$ through Bayesian inference theory. Here, the parameter $x$ can be the coupling strength $g$ or the dimerization parameter $\delta$. 
(b) Energy bands and BSs' properties. 
When QE's frequency is tuned to lie in the middle band gap with width $4|\delta|$, 
two exponentially localized BSs appear, whose energies and wavefunction amplitudes are shown by squares and bars, respectively. 
Here, we choose the transition frequency $\Delta=0$, the coupling strength $g=0.1J$, the dimerization parameter $\delta=0.2$, and the system size $N=41$. }
\label{fig:Fig_1}
\end{figure}

Considerable efforts have been made to achieve quantum-enhanced sensing, most focusing on highly entangled Greenberger-Horne-Zeilinger \cite{Lloyd2004Science, Lloyd2006PRL, Frowis2011PRL} or N00N \cite{Jonathan2008, Joo2011PRL,PhysRevLett.115.250502} states, spin-squeezed states \cite{Wineland1992PRA, Kitagawa1993PRA}, and ground states of many-body systems at the critical point \cite{Zanardi2008PRA,PhysRevLett.128.093401,LIANG20222550,Liang:23,PhysRevLett.117.275301}. 
Although these states have their advantages, they suffer from several challenges, such as preparation and manipulation in experiments \cite{Demkowicz2012NC, Zakrzewski2018PRX,Zhao2021,McConnell2015}. 
Therefore, it is urgent to propose other kinds of quantum sensing protocols, which are easy to implement in experiments, to enhance quantum metrology~\cite{Ginzburg2016Cavity, Kadochkin2018Quantum, Kislov2020Diffusion}. One particular interesting protocol is sensing with a single qubit \cite{Kolkowitz2012Science, Mamin2013Science, Lukin2015Science, Bonato2016NN, Clemmen2016PRL, Poggiali2018PRX, Nakamura2020Science, Kornher2020PRL, Jackson2021NP, Wang2022PRX}, 
and it has been realized in a variety of experimental platforms, including nitrogen-vacancy center \cite{Kolkowitz2012Science, Mamin2013Science, Lukin2015Science, Bonato2016NN}, rare-earth electron \cite{Kornher2020PRL} and nuclear \cite{Jackson2021NP} spin qubit, photonic \cite{Clemmen2016PRL} and superconducting \cite{Nakamura2020Science} qubit. 
  
In this work, we introduce and characterize a novel protocol for sensing with a single qubit, which is inspired by the topological waveguide quantum electrodynamics (QED) \cite{Barik2018Science, Cirac2019SA, Kim2021PRX, Leonforte2021PRL, Bernardis2021PRL, Vega2021PRA, Gong2022PRL, Cheng2022PRA, Bello2022PRXQuantum, Vega2023PRR, Bello2023PRB, Kvande2023PRA, Sheremet2023RMP, Joshi2023PRX, Anderson2016PRX, Tabares2023PRL}. The setup is shown in Fig.~\ref{fig:Fig_1} (a), where a quantum emitter (QE) is coupled to topological baths. The key role of quantum sensing for this setup is the two topological-paired bound states (BSs), which are localized around the QE when the QE’s transition frequency lies deep in the middle band gap \cite{Vladimir1975, John1990PRL, Kurizki1990PRA}. Inspired by this picture, we then systematically investigate the properties of a two-level QE coupled to a topological Su-Schrieffer-Heeger (SSH) waveguide. 
We find that when two BSs appear, the result of projective measurement on the QE can indeed be utilized for achieving quantum-enhanced sensitivity, and reach the Heisenberg limit with coherence time as a quantum resource \cite{Braun2018RMP}. Furthermore, this Heisenberg limit can be extracted based on the Bayes parameter estimation. More importantly, we note that the sensitivity can be maintained across a large range of parameters rather than the critical point. And inheriting from the bath's topological properties, our sensing protocol is robust to disorders of the bath, fluctuations of parameters, and other imperfections.   
  
\section{Quantum Sensing Protocol}

\subsection{Model}

The system that we consider is shown in Fig.~\ref{fig:Fig_1} (a): A two-level QE interacts with a SSH chain that is characterized by having alternating hopping amplitudes $J_{\pm}=J(1\pm\delta)$ between two interspersed sublattices $A$ and $B$. The QE couples to sublattice $A$ at the leftmost site with a coupling strength $g$, and the transition frequency between the excited state $\ket{1}$ and the ground state $\ket{0}$ is $\Delta$. 
The total Hamiltonian of the system reads $\hat{H}=\hat{H}_{\rm E}+\hat{H}_{\rm S}+\hat{H}_{\rm I}$ with \cite{Cirac2019SA}
\begin{subequations}
\label{eq:Ham}
\begin{align}
\label{eq:HE}
\hat{H}_{\rm E} &= \Delta\hat{\sigma}_{+}\hat{\sigma}_{-}\, ,  \\
\label{eq:HS}
\hat{H}_{\rm S} &=  -\sum^{N-1}_{n=1}J_{n}\rbraket{\hat{a}^\dag_n \hat{a}_{n+1} + {\rm H.c.}}\, ,  \\
\label{eq:HI}
\hat{H}_{\rm I} &= g\rbraket{\hat{a}^\dag_1\hat{\sigma}_{-} + \hat{a}_1\hat{\sigma}_{+}} \, .
\end{align}
\end{subequations}
Here, $N$ is the system size, $J_n=J_-$ ($J_+$) for $n$ odd (even), and $\hat{a}^\dag_n$ ($\hat{a}_n$) is the creation (annihilation) operator for a bosonic mode at the $n$-th site of the lattice. $\hat{\sigma}_{-}=\rbraket{\hat{\sigma}_{+}}^\dag=\ket{0}\!\bra{1}$ are the usual pseudospin ladder operators of the QE. 

The Hamiltonian $\hat{H}_\text{S}$ in Eq.~\eqref{eq:HS} describes the topological SSH chain. Assuming periodic boundary conditions, $\hat{H}_\text{S}$ can be rewritten as $\hat{H}_\text{S} = \sum_k f(k) \hat{a}_k^\dagger \hat{b}_k + \text{H.c.}$ in the momentum space. Here, $f(k) = - (J_- + J_+ e^{- i k})$, $\hat{a}_k$ ($\hat{b}_k$) are the Fourier components of $\hat{a}^\dag_n$ for odd (even) $n$. Then the Hamiltonian can be easily diagonalized, which leads to two bands with energy $\omega_{\pm}(k) = \pm J \sqrt{2 (1 + \delta^2) + 2 (1 - \delta^2) \cos(k)}$ and the band gap is thus $4 |\delta| J$. Moreover, both bands can be characterized by a topological winding number $\mathcal{W}$, such that $\mathcal{W} \neq 0$ corresponds to a nontrivial insulator \cite{Chiu2016Classification}.

For open boundary conditions, due to bulk-edge correspondence \cite{Chiu2016Classification} and Lieb theorem \cite{Lieb1989PRL}, the dimerized SSH chain always has a zero energy mode (ZEM) for $\forall \delta\neq0$ if $N$ is odd. Specifically, this ZEM will localize at the left (right) boundary if $\delta > 0$ ($\delta < 0$). 
However, for even $N$, no ZEM exists if $\delta < 0$ while two localized ZEMs will respectively appear at each end of the SSH chain if $\delta > 0$. 
More interestingly, we here find that the system described by Eq.~\eqref{eq:Ham} gives rise to two localized BSs under conditions: I) the SSH bath chain has a ZEM localized near the QE and II) the QE's frequency is tuned to lie in the middle band gap of the bath dispersion, as shown in Fig.~\ref{fig:Fig_1} (b).
This result can be first qualitatively understood based on the degenerate perturbation theory. If we set the frequency $\Delta=0$ and the coupling strength $g=0$, there are two degenerate zero energy states, i.e., $\ket{\Phi_0}=\ket{0}\otimes\ket{\phi_0}$ and $\ket{\Phi_1}=\ket{1}\otimes\ket{\phi_0}$, where $\ket{\phi_0}$ denotes the ZEM of the SSH chain. 
Then we tune on a small coupling strength $g \ll J$, with zeroth order approximation, these two eigenstates are corrected to be $\ket{\Phi_{{\rm B},\pm}}=\frac{1}{\sqrt{2}}\rbraket{\ket{\Phi_0}\pm\ket{\Phi_1}}$, and can thus be qualitatively viewed as the BSs. 

In fact, we can exactly solve the Schr\"odinger equation to obtain these BSs.
Without loss of generality, we consider the resonance case $\Delta=0$ and the SSH chain with an odd number of sites and $\delta > 0$ such that a ZEM exists at the leftmost site (See Appendix B for the case of even $N$). 
With Siegert boundary conditions, the two BSs $\ket{\Phi_{{\rm B},\pm}}$ can be analytically obtained by solving a series of coupling equations (See Appendix A for details). 
For these two BSs $\ket{\Phi_{{\rm B},\pm}}$, the corresponding eigenenergies are $E_{{\rm B},\pm}=\pm E_{\rm B}$ and the overlaps with the state that the QE is excited are $b_\pm=\bra{\Phi_{{\rm B},\pm}}\rbraket{\ket{1}\otimes\ket{\rm vac}} = \mp b$. Here, $|\text{vac}\rangle$ denotes the vacuum state of the SSH chain \cite{Cirac2019SA}, 
\begin{subequations}
\label{eq:BS}
\begin{align}
\label{eq:E_B}
E_{\rm B} &= g\,\sqrt{1+\frac{J^2_-}{g^2-J^2_+}} \, , \\
\label{eq:b}
b &=  \frac{1}{\sqrt{1+\frac{1}{g^2(1-q^2)}\sbraket{E^2_{\rm B}+\rbraket{\frac{g^2-E^2_{\rm B}}{J_-}}^2}}} \, ,
\end{align}
\end{subequations}
with $q = J_-J_+/\rbraket{g^2-J^2_+}$, and we have assumed $|g|<2J|\delta|$ such that the QE's frequency lies in the middle band gap (See Appendix A for details). 
In Fig.~\ref{fig:Fig_1} (b), we plot the wavefunction amplitudes of the two BSs and their corresponding energies. 
We see that these two BSs are exponentially localized at the left boundary and have opposite energies, which are the key to our quantum-enhanced sensing protocol as described below.

\begin{figure}[htp!]
  \centering
  \includegraphics[width=0.48\textwidth]{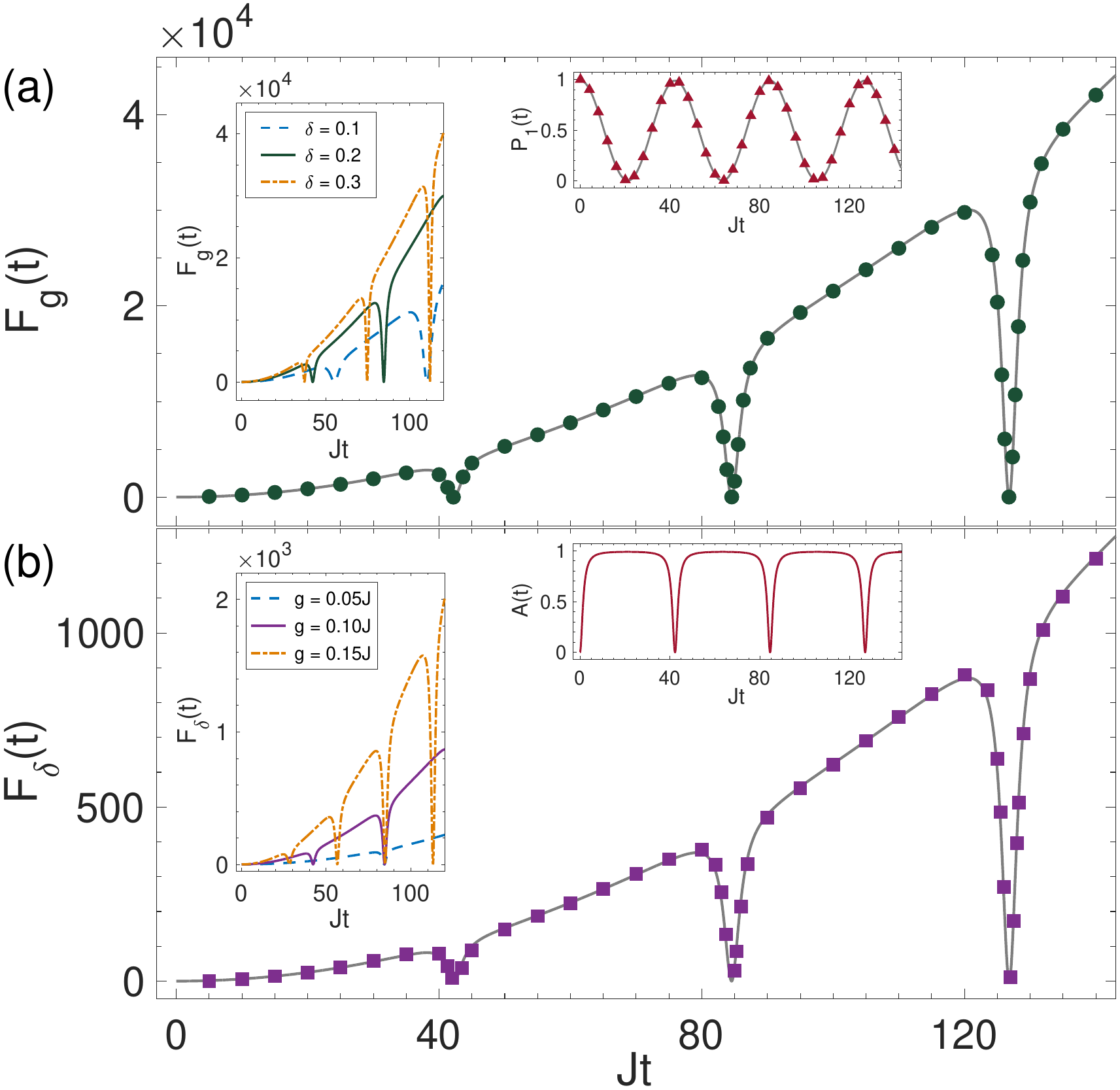}
  \caption{
  Classical Fisher information (a) $F_g(t)$ and (b) $F_\delta(t)$ are plotted as green dots and purple squares, which can be precisely approximated by $\widetilde{F}_g(t)$ and $\widetilde{F}_\delta(t)$ (gray lines), respectively. 
  In the right inset of (a), QE's excited state occupation probability $P_1(t)$ is shown by red triangles, which can be well approximated by $\widetilde{P}_1(t)$ (gray line). 
  The right inset of (b) shows the factor $A(t)$ appeared in Eq.~\eqref{eq:F_x}, which has periodical dips. 
  Here, we choose the transition frequency $\Delta=0$, the coupling strength $g=0.1J$, the dimerization parameter $\delta=0.2$, and the system size $N=201$. 
  Besides, the left insets of (a) and (b) show $F_g(t)$ ($g=0.1J$) and $F_\delta(t)$ ($\delta=0.2$) as functions of $t$ for different $\delta$ and $g$, respectively. 
  }
  \label{fig:Fig_2}
\end{figure}

\subsection{Sensing precision bounds}

To achieve high discrimination of the coupling strength $g$ or the dimerization parameter $\delta$, we monitor the dynamics of the topological waveguide QED system. 
Through the quantum dynamics, the information of $g$ and $\delta$ are transferred to the excitation population of QE which can be subsequently read out [see Fig.~\ref{fig:Fig_1} (a) for a pictorial illustration]. 
To be specific, we consider the unitary evolution $\ket{\psi(t)}=\exp(-i\hat{H}t)\ket{\psi(0)}$ starting from an excited QE $\ket{\psi(0)}=\ket{1}\otimes\ket{\rm vac}$, an initial state that can be easily prepared in experiments \cite{Kim2021PRX}. 
Then the survival amplitude of the initial state in the evolving state is $S(t)=\braket{\psi(0)|\psi(t)}$.

The sensing precision for estimating an unknown parameter $x$ is limited by the celebrated Cram\'er-Rao bound, i.e., the variance ${\rm Var}(x_{\rm est})\geq1/F_x$, where $x_{\rm est}$ is an estimator of $x$ and $F_x$ is Fisher information \cite{Rao1992, Cramer1999}. 
Here, we estimate the coupling strength $g$ or the dimerization parameter $\delta$ by measuring the occupation probability of the QE. The probability of finding the QE in the excited state is $P_1(t)=|S(t)|^2$. 
Therefore, the corresponding sensing bound is given by the classical Fisher information $F_x(t) = [\partial P_1(t)/\partial x]^2/(P_1(t)[1-P_1(t)])$ \cite{Pezze2018RMP}, where $x=g,\,\delta$. 
As shown in Fig.~\ref{fig:Fig_2}, the numerical results for $F_g(t)$ and $F_\delta(t)$ are plotted as green dots and purple squares, respectively. 
Except for periodic dips, we see that both of them have parabolic envelopes, implying the Heisenberg scaling. 
In the following, we will see that these properties of the Fisher information can be well characterized with reasonable approximation.  

We first consider the spectrum decomposition of the Hamiltonian $\hat{H}=\sum_l E_l\ket{\Phi_l}\!\bra{\Phi_l}$, from which we have the survival amplitude $S(t)=\sum_l|C_l|^2\exp(-iE_lt)$, where $C_l=\braket{\Phi_l|\psi(0)}$. 
Except for the two BSs $\ket{\Phi_{B,\pm}}$, the rest of the eigenstates of $\hat{H}$ are extended bulk states, whose overlaps with the initial state can be safely nelected. Therefore, the survival amplitude $S(t)$ can be well approximated by 
\be
\widetilde{S}(t) = \sum_{\alpha=\pm}|\!\braket{\Phi_{B,\alpha}|\psi(0)}\!|^2 e^{-iE_{B,\alpha}t}
=2b^2\cos(E_Bt)\, .
\ee
The probability $P_1(t)=|S(t)|^2$ can thus be precisely approximated by $\widetilde{P}_1(t) = |\widetilde{S}(t)|^2 = 4b^4\cos^2(E_Bt)$ [see the right inset of Fig.~\ref{fig:Fig_2} (a)]. 
Substituting $P_1(t)$ by $\widetilde{P}_1(t)$ and keeping only the quadratic time dependent term, $F_x(t)$ can be approximated by 
\be
\widetilde{F}_x(t) = 
4\rbraket{\frac{\partial E_B}{\partial x}}^2A(t)\, t^2 \, .
\label{eq:F_x}
\ee
Here, the factor $A(t)=4b^4\sin^2(E_b t)/[1-4b^4\cos^2(E_b t)]$, which is periodically time-dependent as illustrated in the right inset of Fig.~\ref{fig:Fig_2} (b). 
Note that these approximations $F_x(t)\approx\widetilde{F}_x(t)$ are extremely accurate as shown in Fig.~\ref{fig:Fig_2}. 
Besides, both $F_g(t)$ and $F_\delta(t)$ are modulated by the periodical time factor $A(t)$, which gives rise to the occurrence of periodical dips as exhibited in Fig.~\ref{fig:Fig_2} (a) and (b). 
However, since this Fisher information has the time-dependent factor $t^2$, we will show in the following that our sensing protocol can achieve the Heisenberg limit with coherence time as quantum resource \cite{Braun2018RMP}. 
Note that although the above approximation is extremely accurate, the system size should be chosen as $N\sim Jt$ such that the Heisenberg limit can be achieved (See Appendix C for details). 
Finally, we see that the dynamics of our topological waveguide QED system can be intuitively understood as Rabi oscillations of the two topological-paired BSs, singling out topological robustness as the key property of our sensing protocol.

Furthermore, in the left inset of Fig.~\ref{fig:Fig_2} (a), we also show $F_g(t)$ for different parameters $\delta$. We find that larger $\delta$ can achieve higher Fisher information since the corresponding BSs are more localized, thus resulting in more susceptibility to the variation of the coupling strength $g$. However, this convenience comes at the cost that more frequent occurrence of the dips. 
A similar phenomenon can be found for $F_\delta(t)$ dependent on the parameter $g$, as illustrated in the left inset of Fig.~\ref{fig:Fig_2} (b). 
These results give the guiding principle for choosing appropriate $\delta$ and $g$ to sense $g$ and $\delta$, respectively.

\begin{figure*}[htp]
\centering
\includegraphics[width=1\textwidth]{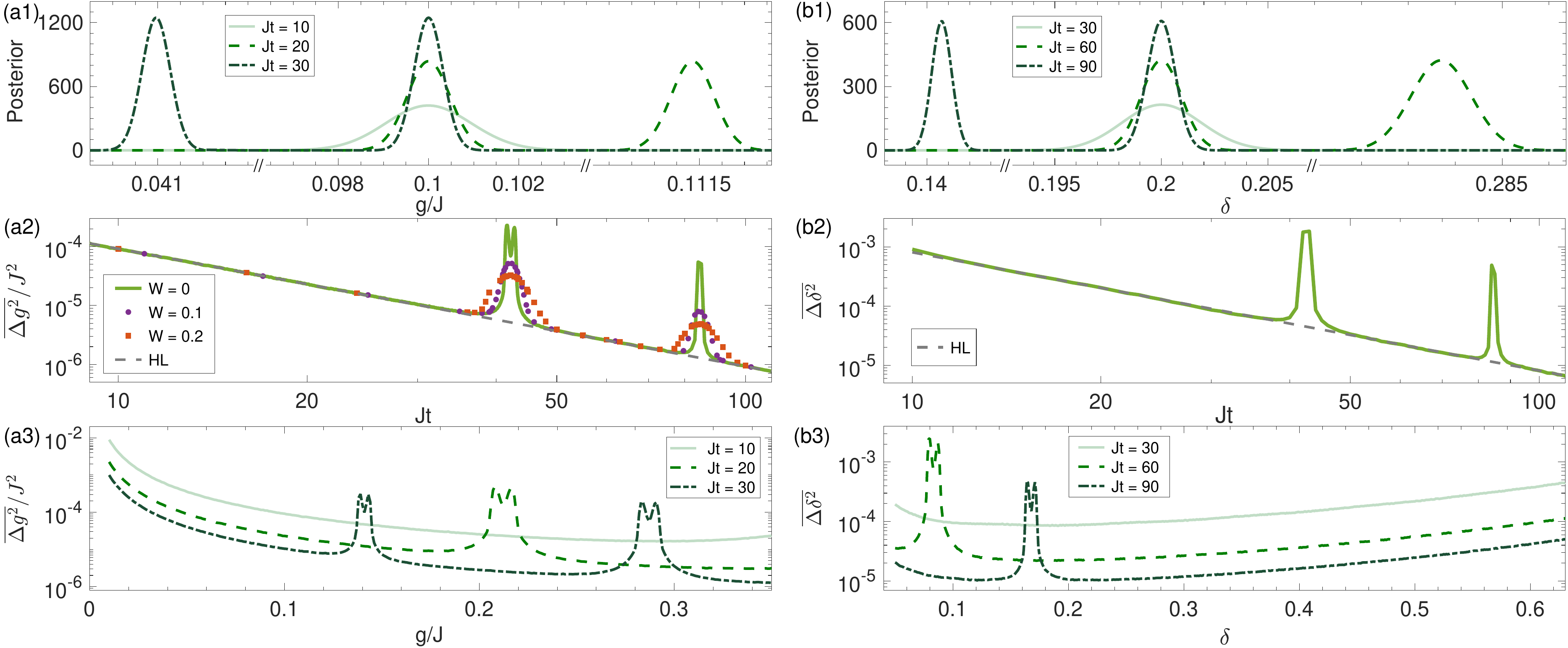}
\caption{
Bayesian parameter estimation for (a) the coupling strength $g$ and (b) the dimerization parameter $\delta$. 
(a1) and (b1), posterior distributions after three different evolution times for sensing $g=0.1J$ and $\delta=0.2$ with uniform priors $g\in[0,\,0.2J]$ and $\delta\in[0,0.4]$, respectively. The Gaussians centered at the true value become narrower by increasing the evolution time. 
(a2) and (b2), average of $\Delta g^2$ and $\Delta \delta^2$ (green solid lines) as functions of $t$ for sensing $g=0.1J$ and $\delta=0.2$, respectively. Both can achieve the Heisenberg limit (HL) indicated by the gray dashed lines. 
In (a2), the purple dots and red squares show the averaged $\Delta g^2$ for disorder strength $W=0.1$ and $W=0.2$, respectively. 
(a3) and (b3), average of $\Delta g^2$ and $\Delta \delta^2$ as functions of $g$ and $\delta$ for three evolution time, respectively. 
Here, we choose the transition frequency $\Delta=0$, the system size $N=201$, and for estimating $g$ ($\delta$), we have fixed $\delta=0.2$ ($g=0.1J$). 
In all plots, the posterior is obtained by repeating the sensing procedure for $M=10^4$ times, and each data point is averaged over $N_{\text{samples}} = 10^4$ samples.
}
\label{fig:Fig_3}
\end{figure*}

\subsection{Bayesian parameter estimation}

Here, we adopt Bayesian parameter estimators to saturate the Cram\'er-Rao bound \cite{Cramer1999, Hradil1996PRL, Pezze2007PRL}, such that the sensing precision can indeed achieve the Heisenberg limit mentioned before. 
The theory of estimation is based on the Bayes rule $P(x | D) = P(D | x ) P(x)/P(D)$. 
Here, the posterior $P(x|D)$ is the conditional probability of the parameter $x$, given the observed data $D$. 
The prior $P(x)$ is the marginal probability of $x$ accounting for the initial information about $x$. 
The prior is updated to the posterior through the likelihood $P(D | x )$, which imprints the measurement results. 
The denominator $P(D)$ is the marginal probability of $D$ accounting for a normalization factor such that $\int{\rm d}x P(x|D)=1$.

For our problem, we repeat the earlier mentioned preparation, evolution, and measurement procedure $M$ times. The probability of finding $m$ outcomes at the excited state of the QE follows the binomial distribution and can be viewed as the likelihood
\be
P(D | x ) = \binom{M}{m}P^m_1(t)[1-P_1(t)]^{M-m}\, .
\ee
Assuming the prior $P(x)$ is a uniform distribution within some interval that we believe the unknown parameter $x$ belongs to, the posterior is then $P(x | D)\propto P(D | x )$. 
As $M$ becomes large, $P(x | D)$ converges to a Gaussian centered at the true value of the unknown parameter $x$ and with a width proportional to the square root of inverse Fisher information \cite{Lehmann1998}. 
To quantify the uncertainty of estimation as well as the bias in the estimation, we adopt the average squared relative error
\be
\Delta x^2 = \frac{\sigma^2_x+|\!\braket{x}-x|^2}{|x|^2}\, ,
\ee
where $\sigma^2_x$ and $\braket{x}$ are the variance and the average of $x$ with respect to the posterior $P(x | D)$, respectively. 
Note that $\Delta x$ reduces to the inverse of the signal-to-noise ratio if the estimation is unbiased. 

Fig.~\ref{fig:Fig_3} shows our numerical results for estimating the coupling strength $g$ and the dimerization parameter $\delta$. 
In Fig.~\ref{fig:Fig_3} (a1), we plot the posterior as a function of $g$ for different evolution times when the true value is $g=0.1J$, where we have assumed a uniform prior $g\in[0,\,0.2J]$ and fixed the dimerization parameter as $\delta=0.2$. 
We see that the Gaussian shapes centered at the true value become narrower by increasing the evolution time, indicating enhancement of the sensing precision. 
On the other hand, since the period of Rabi oscillations, some other Gaussian shapes can appear away from the true value. This ambiguity, however, can be solved by comparing measurements with slightly different evolution times. 
Our sensing protocol can indeed achieve the Heisenberg limit as shown in Fig.~\ref{fig:Fig_3} (a2), where the average squared relative error $\overline{\Delta g^2}\sim t^{-2}$ except for some peaks. 
We have checked that these peaks result from the dips in the corresponding Fisher information, and can be easily avoided by choosing other evolution times. 
To further show the generality of our sensing protocol, in Fig.~\ref{fig:Fig_3} (a3), we plot $\overline{\Delta g^2}$ over a broad range of $g$ for different evolution times. 
As evidenced by the results, increasing coherence time can significantly enhance the sensing precision across a wide range of $g$, except for those regions around the peaks, which again result from the dips in the corresponding Fisher information as we have checked. 
Similar results can be found in Fig.~\ref{fig:Fig_3} (b1-b3) for estimating the dimerization parameter $\delta$, where we have fixed the coupling strength as $g=0.1J$. 

Another remarkable property of our sensing protocol is the robustness to disorder. 
Here, we consider the earlier mentioned resonance condition $\Delta=0$, which has been widely used in the topological waveguide QED system \cite{Cirac2019SA}. 
Under this condition, our model Hamiltonian $\hat{H}$ has chiral symmetry $\hat{C}\hat{H}\hat{C}=-\hat{H}$ with $\hat{C} = \ket{1}\!\bra{1}+\sum^{N}_{n=1}(-1)^n\hat{a}^\dag_{n} \hat{a}_{n}$. Inheriting from this, the BSs are robust to disorder and so it is when sensing the coupling strength $g$. 
To illustrate it, we study the effect of off-diagonal disorder, which appears in the hopping
amplitudes between sublattices $A$ and $B$. 
In this case, the bath's Hamiltonian becomes $H_S\to H_S+J\sum_n(w_n\hat{a}^\dag_{n} \hat{a}_{n+1}+{\rm H.c.})$, where we take the disorder $w_n$ from a uniform distribution within the range $[-W/2,W/2]$. 
As shown in Fig.~\ref{fig:Fig_3} (a2), the disorder diminishes and broadens the peaks of $\overline{\Delta g^2}$. Other than that, it barely affects the precision in estimating $g$ within a considerable range of disorder strength. 
In addition, the dephasing effect has also been considered, and the quantum-enhanced sensitivity can also be achieved even when the dephasing rate is up to $ 5\% J$, which indicates that our sensing protocol is robust against such dephasing noise (See Appendix D for details). 

\begin{figure}[htb]
  \includegraphics[width=0.48\textwidth]{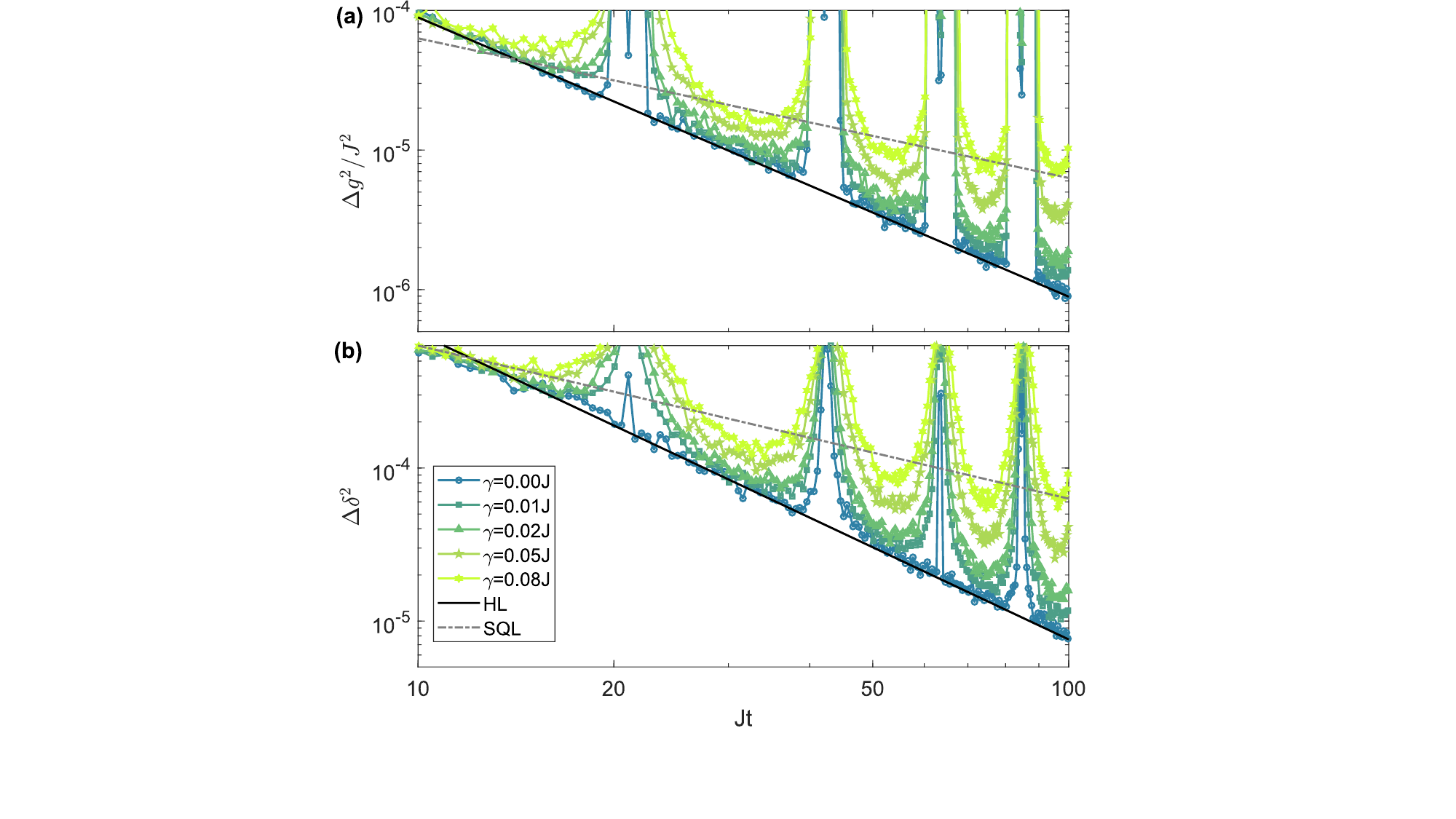}
  \caption{
  Average squared relative errors of Bayesian parameter estimation for (a) the coupling strength $g$ and (b) the dimerization parameter $\delta$. 
  Here, for sensing $g=0.1J$ ($\delta=0.2$), we have fixed $\delta=0.2$ ($g=0.1J$). 
  The system size and the transition frequency are chosen as $N=201$ and $\Delta=0$, respectively. 
  For both panels, the posterior is obtained by repeating the sensing procedure for $M = 10^4$ times, and each data point is averaged over $10^2$ samples. 
  Besides, we also show the standard quantum limit (SQL) using the black dashed line and Heisenberg limit (HL) using the black solid line in both panels. 
  }
  \label{fig:6}
\end{figure}

\subsection{Experimental realization}

We have shown that our sensing protocol can achieve the Heisenberg limit with coherence time as a quantum resource. However, the dephasing effect is inevitable in real experiments and may degrade the sensing precision. 
We here consider that the quantum dynamics of our system undergo a dephasing process for the QE with a constant rate $\gamma$, governed by the master equation 
\bea
    \dot{\rho}(t) =
    \! &-& \! i[\hat{H}, \rho(t)] \nn \\
    \! &+& \!\gamma\sbraket{\left|1\right\rangle\!\left\langle 1\right|\rho(t)\left|1\right\rangle\!\left\langle 1\right|-\frac{1}{2}\{\left|1\right\rangle\!\left\langle 1\right|,\rho(t)\}},
\eea
where $\hat{H}$ is the system's Hamiltonin [see Eq.~\eqref{eq:Ham}] and $\rho$ is the system's density matrix. 
Now the probability of finding the QE in the excited state is $P_1(t) = \Tr{\rho(t)\ket{1}\!\bra{1}}$, and from which we can estimate the coupling strength $g$ and the dimerization parameter $\delta$ with the Bayesian method as mentioned earlier. 
Our numerical results are shown in Fig.~\ref{fig:6}, and we see that as the dephasing rate $\gamma$ increases, the sensing precisions for both $g$ and $\delta$ become worse. 
However, the quantum-enhanced sensitivity can still be achieved till the dephasing rate is up to $\gamma \sim 8\% J$.

In experiments, the dephasing rate is usually much smaller than the coupling strength $J$, which can be achieved by using superconducting qubits \cite{Kim2021PRX} and Rydberg arrays \cite{Browaeys2019Science}.
In superconducting qubits~\cite{Kim2021PRX}, photonic lattice sites are physically realized as $LC$ resonators. There the intracell and intercell coupling capacitances exclusively connect adjacent resonators, thereby strictly enforcing nearest-neighbor interactions. The quantum emitter is implemented through a superconducting transmon qubit integrated into this architecture. With their experimental parameters, the corresponding parameters in the SSH model to be $J \sim 2 \pi \times 356$ MHz and $\gamma \sim 2 \pi \times 0.25$ MHz $ \sim 0.07\% J$ that is much smaller than the strength we considered. In Rydberg arrays~\cite{Browaeys2019Science}, the lattice sites are realized as Rydberg atoms and the hopping mechanism originates from dipolar exchange interactions. The corresponding parameters in the SSH model to be $J \sim 2 \pi \times 1$ MHz and $\gamma \sim 2 \pi \times 1$ KHz $\sim 0.1\% J$ that is still smaller than the strength we considered. Therefore, our sensing protocol is robust against such dephasing noise and it can be safely implemented in these platforms.

\section{Conclusion}

We have devised a versatile and robust protocol for quantum sensing with the topological waveguide QED system, which can achieve the Heisenberg limit precision across a large range of parameters. 
Through analytical investigation, we show the key to our protocol is the two paired BSs that inherit the topological robustness of the bath. 
Our sensing protocol thus paves the way for the development of topological quantum sensors, which are expected to be robust against local perturbations. 
Besides, the single-qubit-based schemes just demand simple projective measurement, thus may soon lead to experimental demonstrations of our protocol. 
Candidate near-term quantum platforms include superconducting qubits \cite{Kim2021PRX} and Rydberg arrays \cite{Browaeys2019Science}. 
Both of them have been utilized to realize the topological SSH chain, and our sensing protocol can be easily implemented in these platforms for their flexible controllability.

\section*{Acknowledgments}
This work is supported by National Key Research and Development Program of China (2021YFA0718303, 2021YFA1400904); National Natural Science Foundation of China (92165203, 61975092, 11974202, 12375023, 12204428); the Natural Science Foundation of Henan Province (242300421159). 
Xingze Qiu acknowledges support from the Fundamental Research Funds for the Central Universities and Shanghai Science and Technology project (24LZ1401600).

\newpage

\renewcommand{\theequation}{S\arabic{equation}}
\renewcommand{\thefigure}{S\arabic{figure}}
\renewcommand{\thetable}{S\arabic{table}}
\setcounter{equation}{0}
\setcounter{figure}{0}
\setcounter{table}{0}

\section*{Appendix A: Analytical Solutions for the Bound States}

In the following, we give analytical solutions for the BSs, which will appear when a ZEM of the SSH chain exists. 
The wave function of the BSs can be parameterized as 
\be
\label{eq:BS_WF}
\ket{\Phi_{\rm B}} = \rbraket{\widetilde{b}\,\hat{\sigma}_+ + \sum^N_{n=1}c_n a^\dag_n}\rbraket{\ket{0}\otimes\ket{\rm vac}}\, .
\ee
$\widetilde{b}$ and $c_n$ are real coefficients. $\ket{\rm vac}$ denotes the vacuum state of the SSH chain. 
Without loss of generality, we assume $0<\delta<1$ and $N$ is odd, such that a ZEM is localized at the left boundary. 
To obtain analytical solutions for the BSs, we adopt the Siegert boundary conditions with outgoing waves in the form \cite{Hatano2008, Hatano2013}
\be
\label{eq:SiegertBC}
\begin{pmatrix}
       c_n \\
       c_{n+1} 
 \end{pmatrix} 
 = \exp\rbraket{ik\frac{n+1}{2}}
 \begin{pmatrix}
       d_1 \\
       d_2
 \end{pmatrix}\, ,
 \quad {\rm for\, odd}\, n\, , 
\ee
where the wave number $k$ is a complex number for the BSs. 
Solving the Schr\"odinger equation $\hat{H}\ket{\Phi_{\rm B}}=\widetilde{E}_{\rm B}\ket{\Phi_{\rm B}}$ with eigenvalue $\widetilde{E}_{\rm B}$ and eigenstate $\ket{\Phi_{\rm B}}$ yields a series of coupled equations as follows
\begin{subequations}
\begin{align}
g\,d_1 &= (\widetilde{E}_{\rm B}-\Delta)\,\widetilde{b}\,e^{-ik} \, , \\
g\,\widetilde{b} &= (\widetilde{E}_{\rm B} d_1+J_-d_2)\,e^{ik} \, , \\
\widetilde{E}_{\rm B}\,d_2 &= -\rbraket{J_- + J_+ e^{ik}}d_1 \, , \\
\widetilde{E}_{\rm B}\,d_1 &= -\rbraket{J_- + J_+ e^{-ik}}d_2\, .
\end{align}
\end{subequations}
Together with the normalization condition $\widetilde{b}^2+\sum^N_{n=1}c^2_n=1$ for the eigenstate $\ket{\Phi_{\rm B}}$ and the resonance condition $\Delta=0$, we have
\begin{subequations}
\label{eq:parameters_sloution}
\begin{align}
\widetilde{E}_{\rm B} &= \pm g\,\sqrt{1+\frac{J^2_-}{g^2-J^2_+}} \, , \\
\widetilde{b} &= \mp \sqrt{1+\frac{1}{g^2(1-q^2)}\sbraket{\widetilde{E}^2_{\rm B}+\rbraket{\frac{g^2-\widetilde{E}^2_{\rm B}}{J_-}}^2}} \, , \\
d_1 &= \frac{1}{qg}\widetilde{E}_{\rm B} \widetilde{b} \, ,\\
d_2 &= \frac{1}{qg}\frac{g^2-\widetilde{E}^2_{\rm B}}{J_-} \widetilde{b}\, , \\
e^{ik} &= q \, ,
\end{align}
\end{subequations}
where $q = J_-J_+/\rbraket{g^2-J^2_+}$, and we have assumed that the system size $N$ is large enough and $|g|<2J|\delta|$. 
By substituting Eq.~\eqref{eq:parameters_sloution} into Eq.~\eqref{eq:SiegertBC} and then Eq.~\eqref{eq:BS_WF}, we obtain the analytical solutions of the wave functions $\ket{\Phi_{\rm B}}$ for the two BSs. Besides, these two BSs are exponentially localized at the left boundary because of $|q|<1$.

\begin{figure}[htb]
    \centering
    \includegraphics[width=0.48\textwidth]{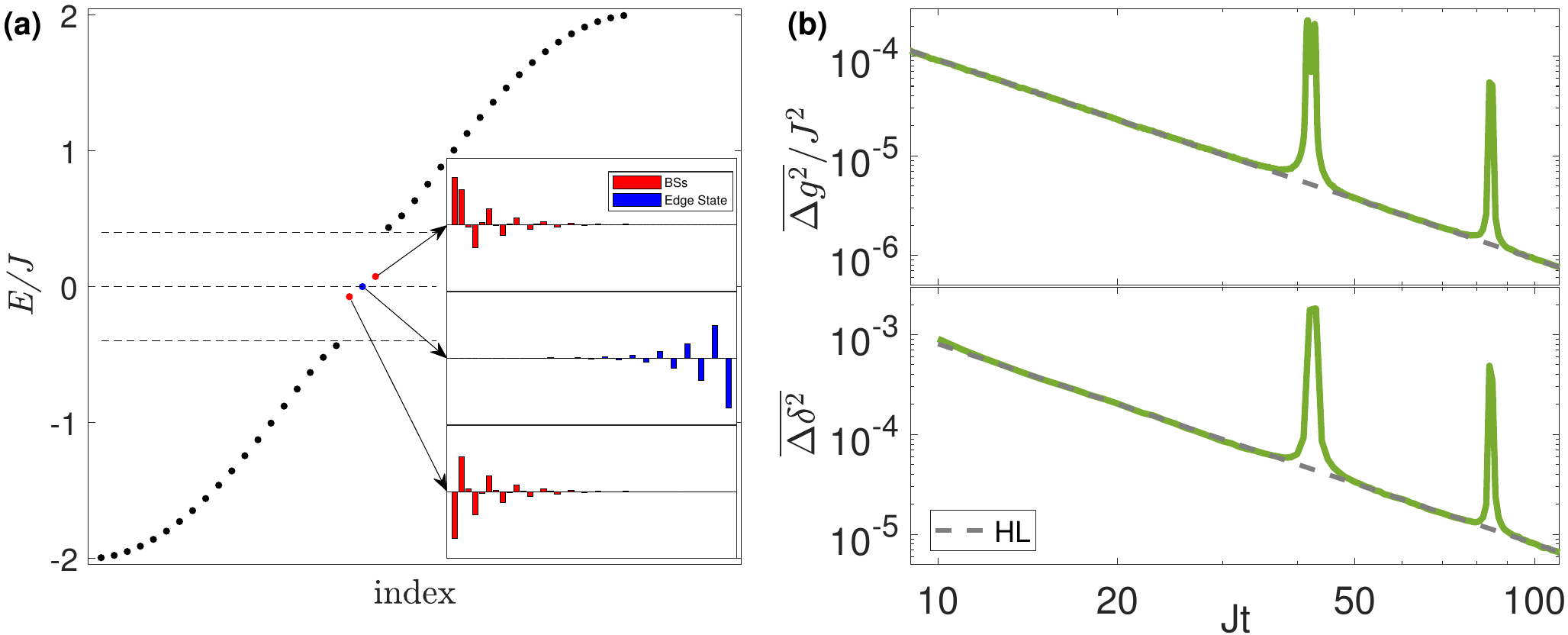}
    \caption{(a) Energy bands, topological-paired BSs, and ZEM. The inset shows wave function amplitudes of the BSs (top and bottom) and the ZEM (middle), respectively. Here, we choose the parameters $N=40$, $\delta = 0.2$ and $g = 0.1J$. 
    (b) Average squared relative errors of Bayesian parameter estimation for the coupling strength $g$ and the dimerization parameter $\delta$. 
    For sensing $g=0.1J$ ($\delta=0.2$), we have fixed $\delta=0.2$ ($g=0.1J$). 
    The system size and the transition frequency are chosen as $N=200$ and $\Delta=0$, respectively. 
    For both panels, the posterior is obtained by repeating the sensing procedure for $M = 10^4$ times, and each data point is averaged over $10^2$ samples. 
    The grey dashed lines indicate the Heisenberg limit (HL). 
    }
    \label{fig:4}
\end{figure}

\section*{Appendix B: Applicability of Our Sensing Protocol for Even System Size}

As discussed in the main text, the number of zero energy modes (ZEMs) in a finite-size Su-Schrieffer-Heeger (SSH) chain depends on the parity of system size $N$ and the sign of $\delta$.
In the main text, we have discussed the situation of odd $N$ and positive $\delta$. 
For odd $N$ and negative $\delta$, our sensing protocol is still applicable by coupling the quantum emitter (QE) to the right end of the chain. 
Here, we numerically discuss the situation of even $N$ and positive $\delta$. In this case, there are two localized ZEMs respectively appearing at each end of the SSH chain. 
As shown in Fig.~\ref{fig:4} (a), when a QE couples to the left-most site of the SSH chain, two topological-paired bound states (BSs) near the QE appear. 
Although there is a ZEM localized at the right end of the chain, this ZEM does not affect the sensing precision since it is far away from the BSs and thus has no effect on the dynamics of the BSs, as illustrated in Fig.~\ref{fig:4} (b).

\section*{Appendix C: Finite Size Effect}

\begin{figure}[htb]
  \centering
  \includegraphics[width=0.48\textwidth]{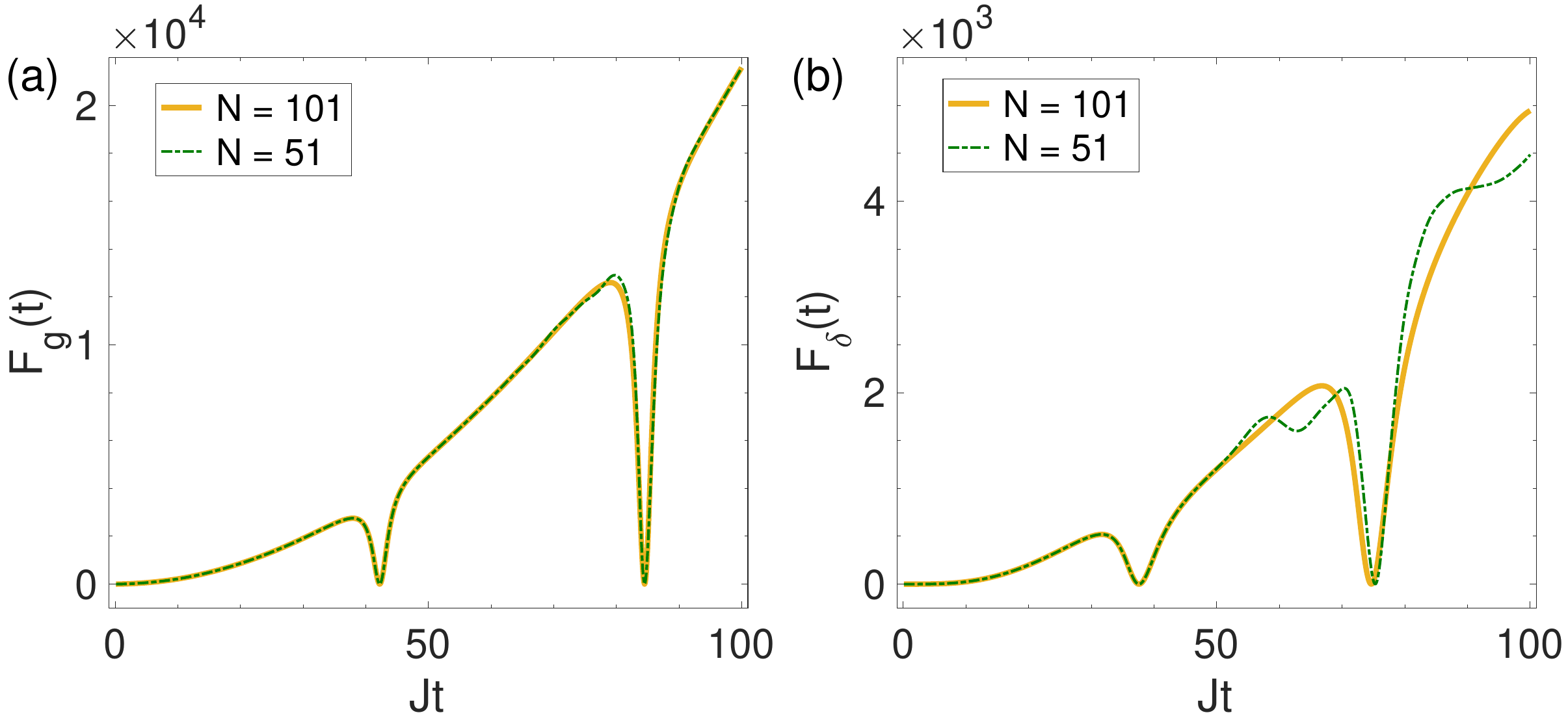}
  \caption{
  Classical Fisher information (a) $F_g(t)$ and (b) $F_\delta(t)$ as a function of coherence time $t$ for different system sizes. 
  Here, we choose the transition frequency $\Delta=0$, the dimerization parameter $\delta=0.2$, and the coupling strength $g=0.1J$.
  }
  \label{fig:5}
\end{figure}

Here, we give more discussion about the requirement that the system size should be chosen as $N \gtrsim  Jt$. 
Despite the very large overlap between the two topological-paired BSs and the initial state, the overlap between the bulk states and the initial state cannot be completely ignored. 
This faintest leakage will spread away from the QE and be rebounded by another end of the SSH chain, and then propagate back towards the QE. 
The total time of this process can be reasonably assumed to be proportional to the system size $N$. 
After this evolution time, the leakage can significantly affect the dynamics of the two-level system formed by the two BSs. 
This is thus named as ``finite size effect". 
As shown in Fig.~\ref{fig:5}, this effect can be directly observed through the classical Fisher information. 
We see that the Fisher information will oscillate when the evolution time $Jt$ is larger than the system size $N$. 
In addition, compared with $F_g(t)$, $F_\delta(t)$ is more susceptible to the finite size effect.

\bibliographystyle{apsrev4-2}
\bibliography{References}

\end{document}